\documentclass[preprint,showpacs,preprintnumbers,amsmath,english,amssymb]{revtex4}
\pdfoutput=1
\usepackage{graphicx}
\usepackage{dcolumn}
\usepackage{bm}

\begin{document}


\title{Magnetization dynamics at elevated temperatures}

\author{Lei Xu}
\author{Shufeng Zhang}%

\affiliation{%
 Department of Physics, University of Arizona, Tucson, AZ 85721, USA
}%

\date{\today}

\begin{abstract}
By using the quantum kinetic approach with the instantaneous local
equilibrium approximation, we propose an equation that is capable of
addressing magnetization dynamics for a wide range of temperatures.
The equation reduces to the Landau-Lifshitz equation at low
temperatures and to the paramagnetic Bloch equation at high
temperatures. Near the Curie temperature, the magnetization reversal
and dynamics depend on both transverse and longitudinal relaxations.
We further include the  stochastic fields in the dynamic equation in
order to take into account fluctuation at high temperatures. Our
proposed equation may be broadly used for modeling laser pump-probe
experiments and heat assisted magnetic recording.
\end{abstract}

\maketitle


\section{Introduction}

The phenomenological Landau-Lifshitz (LL) equation is the basis of
powerful micromagnetic codes for simulation of magnetic structure
and dynamics in magnetic materials. The key component of the LL
equation is that magnetization relaxation during dynamic processes
is described by a single damping parameter $\alpha$ \cite{Landau},
\begin{equation}\label{eq:rashba}
\frac{d \bf m}{dt} = -\gamma {\bf m}\times {\bf H}_{\rm eff} -
\gamma \frac{\alpha}{m} {\bf m}\times ({\bf m}\times {\bf H}_{\rm
eff})
\end{equation}
where ${\bf m}({\bf r},t)$ is the magnetization density vector which
is a function of space and time, $m=|{\bf m}|$ is its magnitude, and
${\bf H}_{\rm eff}$ is the effective magnetic field including the
magnetic anisotropic, magnetostatic and external fields. The second
term on the right side of the equation describes a phenomenological
{\em transverse} relaxation since the magnitude of the magnetization
density $m$ is conserved. Such transverse relaxation model is indeed
a valid approximation because the magnetization $m$ (the order
parameter) of the ferromagnet is nearly independent of the magnetic
field as long as the temperature is not too close to the Curie
temperature.

Recently, there is an emerging technological need to extend the LL
equation to high temperatures in order to model the dynamics near or
above the Curie temperature for laser-induced demagnetization (LID)
\cite{Bigot,Rasing} and heat-assisted magnetic recording (HAMR)
\cite{Kryder,Naturetech}. Due to the strong fluctuation of the
magnetic momentum, one needs to reduce the size of magnetic cells if
one continues to use the LL equation to model the magnetization
dynamics. When the cell size reduces to the ultimate smallest size
of magnetic atoms, the so-called atomistic LL equation which has the
same form as the conventional LL equation \cite{Garanin} had been
proposed,
\begin{equation}
\frac{d{\bf S}_i}{dt} =-\gamma {\bf S}_i \times {\bf H} - \gamma
\alpha {\bf S}_i \times ({\bf S}_i \times {\bf H})
\end{equation}
where ${\bf H}$ is an effective magnetic field (treated as a
c-number) including a random fluctuating field and ${\bf S}_i$ is
the spin of $i$th atom which is treated classically. While the above
atomistic LL equation might qualitatively capture some of static and
dynamic properties near the Curie temperature \cite{Chuby,Atx}, we
point out below that the above atomistic LL equation has several
fundamental problems.

First, the spin $S_i$ in ferromagnetic metals such as Ni, Co, Fe and
their alloys is usually small. The replacement of the quantum spin
by the classical vector severely neglects the quantum nature of the
spin fluctuation of atomic spins. More importantly, the atomistic LL
equation, Eq.~(2), has no microscopic origin and it is fundamentally
incompatible with quantum mechanics. For example, if one takes the
case for $S_i=1/2$ (e.g., Ni). The second ("damping") term of
Eq.~(2) becomes $\gamma \alpha (\frac{i}{2}{\bf S}_i \times {\bf H}
-\frac{1}{2} {\bf H})$ for a quantum spin and thus Eq.~(2) becomes
\begin{equation}
\frac{d{\bf S}_i}{dt} =-\gamma (1+i\alpha/2){\bf S}_i \times {\bf H}
+\gamma \alpha {\bf H}/2.
\end{equation}
This unphysical equation originates from the broken time-reversal
symmetry inherited on the atomistic LL equation.

The second difficulty is that the atomistic LL equation is not
derivable from an effective Hamiltonian, even at the
phenomenological level. If the atomistic LL equation has some
validity, a microscopic or an effective Hamiltonian should exist.
For example, if we construct a spin Hamiltonian of the form $H'
\propto \sum_i {\bf S}_i \cdot ({\bf H} + \alpha {\bf S}_i\times
{\bf H})$, the equation of motion for ${\bf S}_i $ would be $d{\bf
S}_i/dt = (1/i\hbar)[{\bf S}_i , H'] $ which results in an
additional term compared to Eq.~(2) due to none-zero commutation
$[{\bf S}_i, \alpha {\bf S}_i \times {\bf H}] \neq 0$. On the other
hand, if one takes the phenomenological Hamiltonian as $H' \propto
\sum_i {\bf S}_i \cdot ({\bf H} + \alpha {\bf m}\times {\bf H}) $
where ${\bf m} = <{\bf S}_i> $ (note that $<>$ denotes ensemble
thermal averaging and thus ${\bf m}$ is a c-number), the result
dynamics for ${\bf S}_i$ would be
\begin{equation}
\frac{d{\bf S}_i}{dt} =\frac{1}{i\hbar} [{\bf S}_i , H'] = -\gamma
{\bf S}_i \times {\bf H} - \gamma \alpha {\bf S}_i \times ({\bf m}
\times {\bf H}).
\end{equation}
The above equation is precisely the original LL equation after the
thermal averaging. Thus, the macroscopic LL equation is derivable
from an effective Hamiltonian while the atomistic LL equation is
not.

In spite of above conceptual difficulties in the atomistic LL
equation, it has been shown that the result derived from the
atomistic LL equation with stochastic fields is in agreement with
the Monte Carlo simulation \cite{Nowak3}. We point out that this
agreement is not surprising: for equilibrium properties such as
magnetic moment and critical exponents, the calculated results are
insensitive to the details of the ``damping''; for dynamic
properties such as reversal time, the Monte Carlo steps are
calibrated to fit the real time in the atomistic stochastic LL
equation \cite{Cheng}. Therefore, such agreement should not be
interpreted as the proof of the validity of the atomistic LL
equation.

In this paper, we propose an effective magnetization dynamic
equation for a wide range of temperatures without assuming the
presence of the atomistic LL equation for each atomic spin. By using
the equation of motion for the quantum density matrix within the
instantaneous local relaxation time approximation \cite{Davies}, we
show that the magnetization dynamics for ferromagnets can be cast in
the form of the Bloch equation for paramagnetic spins \cite{Bloch}.
In Sec.II, we explicitly derive the generalized Bloch equation and
show that the equation is consistent with the known dynamics at low
and high temperatures. In Sec. III, we analyze the longitudinal and
transverse relaxations from our result, and apply our effective
equation to study the magnetization reversal processes near Curie
temperatures. Finally, we add necessary stochastic fields in the
equation to capture the fluctuation of the dynamics.

\section{Effective dynamic equation for ferromagnets}

We start with a density operator $\hat{\rho}$ which may be written
in the spinor form $\hat{\rho}=\rho_1 + \mbox{\boldmath $\sigma$}
\cdot \mbox{\boldmath $\rho$}_2$ where $\rho_1$ and $\mbox{\boldmath
$\rho$}_2 $ are spin-independent and spin-dependent density
operators, and $\mbox{\boldmath $\sigma$}$ is the Pauli matrix
vector. Within the instantaneous local relaxation time
approximation, the density operator satisfies the quantum kinetic
equation \cite{Davies}
\begin{equation}
\frac{d\hat{\rho}}{dt} = \frac{1}{i\hbar} [\hat{\rho}, \hat{H}] -
\frac{\rho_1 - \bar{\rho}_1}{\tau_p} - \mbox{\boldmath
$\sigma$}\cdot \frac{\mbox{\boldmath $\rho$}_2 - \mbox{\boldmath
$\bar{\rho}$}_2}{\tau_s}
\end{equation}
where $\bar{\rho}_1$ and $\mbox{\boldmath $\bar{\rho}$}_2$ are the
{\em instantaneous local equilibrium} (ILE) densities; they are
different from the static equilibrium values. In electron transport
theories, these ILE densities depend on the local chemical potential
$\mu ({\bf r})$ or the local electric field ${\bf E}({\bf r},t)$ and
they are in turn related to the densities themselves. For example,
for spin dependent electron transport, the inclusion of the spin
relaxation (third term of Eq.~5) leads to the well-known
spin-diffusion equation for the spin dependent chemical potential
(or spin density) \cite{Valet}. In the present case, these ILE
densities are functions of the local effective magnetic field. At a
given time, the effective field consists of the ferromagnetic
exchange, anisotropy, external, and classical magnetostatic field;
we will discuss these fields in more details later. The two
relaxation times $\tau_p$ and $\tau_s$ represent the momentum and
spin relaxation times; these two relaxation times control the
electron charge diffusion (conductance) and spin diffusion
(spin-dependent transport). If we now consider an effective
Hamiltonian $\hat{H}=\hat{H}_0 - g\mu \mbox{\boldmath $\sigma$}\cdot
{\bf H}_{\rm t}(t) $ where $\mu$ is the Bohr magneton, $\hat{H}_0$
is treated as an unperturbed Hamiltonian, we find the
self-consistent equation for the magnetization ${\bf m}\equiv g\mu
{\rm Tr}(\mbox{\boldmath $\sigma$}\hat{\rho})= g\mu {\rm Tr}
\mbox{\boldmath $\rho$}_2$ readily from Eq.~(5),
\begin{equation}
\frac{d{\bf m}}{dt} =-\gamma {\bf m}\times {\bf H}_{\rm t} -
\frac{{\bf m}-{\bf m}_{eq} ({\bf H}_{\rm t})}{\tau_s}.
\end{equation}
where the ILE magnetization ${\bf m}_{\rm eq} = g\mu \mbox{\boldmath
$\bar{\rho}$}_2$ is identified as the thermal equilibrium value for
a given magnetic field ${\bf H}_{\rm t}$.

At the first sight, Eq.~(6) is similar to the well known Bloch
equation \cite{Bloch} that has been widely used for understanding
nuclear spin resonance experiments. In the Bloch equation, the
equilibrium magnetization ${\bf m}_{\rm eq}$ is a known equilibrium
state which is related to the dynamic susceptibility $\chi ( \omega
) $, i.e., ${\bf m}_{eq} = \chi {\bf H}_{\rm ext}$ and ${\bf
m}_{eq}$ is independent of ${\bf m} (t)$. In the present content,
${\bf m}_{eq}$ is not known {\em a priori} and ${\bf m}_{eq}$ varies
with time. At any time $t$, there is an instantaneous equilibrium
magnetization ${\bf m}_{\rm eq}$ that depends on the total magnetic
field ${\bf H}_t$. To solve Eq.~(6), one first needs to model the
instantaneous local field ${\bf H}_{\rm t}$ and its relation to
${\bf m}_{\rm eq}$.

In the conventional LL equation, the effective field ${\bf H}_{\rm
eff}$ consists of the external field, the anisotropy and the
magnetostatic (dipole) fields. The exchange field which comes from
the ferromagnetic exchange interaction between neighboring spins is
included only when there is spatial variation in the magnetic domain
structure. The uniform exchange term, $J{\bf m}$, is unimportant
since it is parallel to the magnetization and it does not contribute
to the LL dynamic equation. In the present case, however, the
exchange interaction is the largest and most important term in
determining the instantaneous equilibrium magnetization ${\bf
m}_{\rm eq}$. We thus model the total instantaneous magnetic field
${\bf H}_t = J{\bf m}+{\bf H}_{\rm eff}$. It is noted that ${\bf
H}_{\rm eff}$ depends on the instantaneous magnetization ${\bf
m}(t)$ as well.

Next, we should establish an explicit relation between the total
field ${\bf H}_t$ with ${\bf m}_{\rm eq}$. There are a number of
approaches available to describe such relation. The simplest
approach would be using the molecular field approximation where the
equilibrium magnetization can be explicitly expressed by
\cite{Aschroft}
\begin{equation}
{\bf m}_{eq} \equiv g\mu <{\bf S}_i> = g\mu S B_S(\beta g\mu H_t)
\hat{\bf H}_t
\end{equation}
where $S$ is the spin of the atom, $\beta =(k_BT)^{-1}$ is the
inverse of temperature, $B_S (x)\equiv (1/S)[(S+1/2)\coth (S+1/2)x
-(1/2)\coth(x/2)]$ is the Brillouin function and $\hat{\bf H}_t$ is
the unit vector in the direction of ${\bf H}_t$, i.e., $\hat{H}_t =
{\bf H}_t/H_t$. In the time-independent case, ${\bf m}={\bf m}_{eq}$
and the above equation is the well-known mean-field result that
determines the ferromagnetic order parameter ${\bf m}_{eq}$. In a
non-equilibrium situation where ${\bf m}$ depends on time, we
interpret ${\bf m}_{eq}$ in Eq.~(7), which is also dependent on
time, as the instantaneous local equilibrium magnetization at a
given (instantaneous) field ${\bf H}_t$.

Our proposed Eq.~(6) supplemented by Eq.~(7) can semi-quantitatively
describe magnetization dynamics at all temperatures. Before we
examine some limiting cases, we comment on certain important
approximations leading to these equations. The instantaneous
relaxation time approximation, Eq.~(5), has been routinely applied
to many quantum or semi-classical systems for transport and magnetic
properties. The accuracy of this approximation is hard to assess for
the ferromagnetic systems. However, the instantaneous relaxation
time approximation has been very successfully applied in spin
diffusion of magnetic multilayers where the semiclassical
distribution function is assumed to relax to the instantaneous
chemical potential \cite{Valet}. Furthermore, the relaxation time
approximation usually serves as a first step in a phenomenological
theory since it gives rise an analytically closed form. The most
severe approximation is to replace ${\bf m}_{\rm eq}$ by the mean
field Brillouin function, Eq.~(7). Such approximations are known to
produce inaccurate critical exponents and Curie temperatures. There
are several much improved approaches such as the renormalization
group theory \cite{x1}, self-consistent random phase approximation
\cite{x2}, and Monte Carlo simulation \cite{x3}. While these
approaches treat the fluctuation near the critical temperature
better, they are far more complicated and without an analytical
form. On the other hand, the mean field approximation is
qualitatively correct and it allows a much simpler description of
magnetization dynamics in spite of underestimating the critical
fluctuation. For the purpose of establishing a phenomenological
dynamic equation similar to the LL equation, we believe that the
choice of the mean field approximation throughout this study is
appropriate.

Similar to the LL equation, Eq.~(6) contains a phenomenological
parameter, $\tau_s$, representing the magnetic relaxation of
paramagnetic spins. In transition metals, $\tau_s$ is related to the
spin-flip time. In fact, there are a number of theoretical and
experimental studies on the numerical values of $\tau_s$ in
different materials \cite{Mertig,Schultz,Wernick}. For transition
metals, the relaxation time ranges from sub-picoseconds to a few
picoseconds.

\section{Longitudinal and transverse magnetization dynamics}

Before we proceed to solve Eq.~(6) in a number of interesting
examples, we examine several limiting cases. First, by using the
identity
\begin{equation}
{\bf H}_t =m^{-2} [({\bf m}\cdot {\bf H}_t) {\bf m} - {\bf m}\times
({\bf m}\times {\bf H}_t) ]
\end{equation}
we write Eq.~(6) in terms of three mutually perpendicular vectors,
\begin{equation}
\frac{d{\bf m}}{dt} =-\gamma {\bf m}\times {\bf H}_{\rm eff}
-\frac{\gamma \alpha_{\rm tr}}{m} {\bf m}\times ({\bf m}\times {\bf
H}_{\rm eff}) - \frac{\gamma \alpha_l}{m} ({\bf m}\cdot {\bf H}_{\rm
t} ) {\bf m}
\end{equation}
where we have introduced the transverse and longitudinal
dimensionless damping coefficients $\alpha_{\rm tr}$ and $\alpha_l$,
\begin{equation}
\alpha_{\rm tr} = \frac{m_{eq}}{\gamma \tau_s m H_t}
\end{equation}
and
\begin{equation}
\alpha_{l} =\frac{1}{\gamma \tau_s }\left[ \frac{m}{{\bf m}\cdot {\bf
H}_t} -\frac{m_{\rm eq}}{mH_t} \right] .
\end{equation}
At low temperatures, ${\bf m}$ is close to $g\mu S$ and the exchange
field $J{\bf m}$ is much larger than the other fields ${\bf H}_{\rm
eff}$. Thus, one immediately has $\alpha_{\rm tr} = (\gamma \tau_s
Jm)^{-1}$. In a typical transition ferromagnet such as Co or Fe, $J$
is of the order of the Curie temperature (0.1-0.2 eV) and $\tau_s$
is a sub-picosecond, we find $\alpha_{\rm tr}$ is of the order of
$10^{-3}-10^{-1}$.

To estimate the low temperature longitudinal relaxation $\alpha_l$
from Eq.~(11), we consider an initial $m$ deviates from the
equilibrium value of $g\mu S B_S$ and from Eq.~(11), $\alpha_l$ would
be about the same order of magnitude as $\alpha_{\rm tr}$. However,
the longitudinal field $J{\bf m}$ is much larger than $H_{\rm eff}$
and thus the ratio of the longitudinal ($\tau_{l}$) to the
transverse ($\tau_{\rm tr}$) relaxation times is about
$\tau_l/\tau_{\rm tr} \approx H_{\rm eff}/J$. Even for a very high
anisotropy material and a large magnetic field, $J$ is several
orders of magnitude larger than $H_{\rm eff}$; this justifies that
at the low temperature one can neglect the longitudinal relaxation
in the dynamic equation, i.e., the magnitude of the magnetization is
always in equilibrium.

When the temperature is much higher than the Curie temperature,
Eq.~(6) represents the paramagnetic Bloch equation. In this case,
the equilibrium magnetization ${\bf m}_{eq}$ may be expressed via
susceptibility $\chi$, i.e., ${\bf m}_{eq} = \chi {\bf H}_{\rm
eff}$. Such dynamic equations have been frequently used for
understanding paramagnetic resonant phenomena where the resonance
width is determined by the relaxation time $\tau_s$.

The most interesting case of Eq.~(6) is for temperature close to
Curie temperature where transverse and longitudinal relaxation times
could become comparable. To see this, we consider the effective
field is parallel to ${\bf m}(t)=m(t) {\bf e}_z$ and expand $B_S (x)
= (S+1)x/3 - (1/90)(S+1)(2S^2+2S+1)x^3$ up to the third order
in the small $x$ where $x=\beta g\mu H_t$. Then, Eq.~(6) for
temperature close to the Curie temperature becomes
\begin{equation}
\frac{dm}{dt} =  - \frac{1}{J \tau_s} \left[ \left( 1-\frac{T_c}{T}
\right) H_t + \frac{3}{10J^2} \frac{T_c^3}{T^3} \left( \frac{1}{S^2} +
\frac{1}{(1+S)^2} \right) H_t^3 - H_{\rm eff} \right]
\end{equation}
where $T_c = S(S+1)J(g\mu)^2/3k_B $ is the mean field Curie
temperature. In the absence of the magnetic field ${\bf H}_{\rm
eff}=0$ and $H_t = Jm$, and we can immediately solve the above
equation,
\begin{equation}
m(t) = m(0) e^{-t/\tau_l} \left[ 1+ G (1-e^{-2t/\tau_l})
\right]^{-1/2}
\end{equation}
where $m(0)$ is the initial magnetization, $G = \frac{3
T_c^3m^2(0)}{10 T^3}\left( \frac{1}{S^2} + \frac{1}{(1+S)^2} \right)
(1-T_c/T)^{-1}$, and
\begin{equation}
\tau_l = \tau_s \left( 1- \frac{T_c}{T} \right)^{-1}.
\end{equation}
Thus, the longitudinal relaxation time, $|\tau_l|,$ near Curie
temperature, is associated with the critical phenomenon. The
relaxation time becomes very long when the temperature approaches
the Curie temperature. The dynamics slow-down at the critical
temperature is in fact a general property of critical phenomena
\cite{critical}. In the presence of the magnetic field, the phase
transition becomes a smooth change and the dynamic slow-down is no
more critical. In Fig.~1, we show the longitudinal relaxation as the
function of the magnetic field and temperature. Clearly, the
magnetic field suppresses the longitudinal dynamic slowdown. It is
noted that the peak of the relaxation time in the presence of the
magnetic field is shifted to higher temperatures.

\begin{figure}
  \centering
  \includegraphics[width=8.5cm]{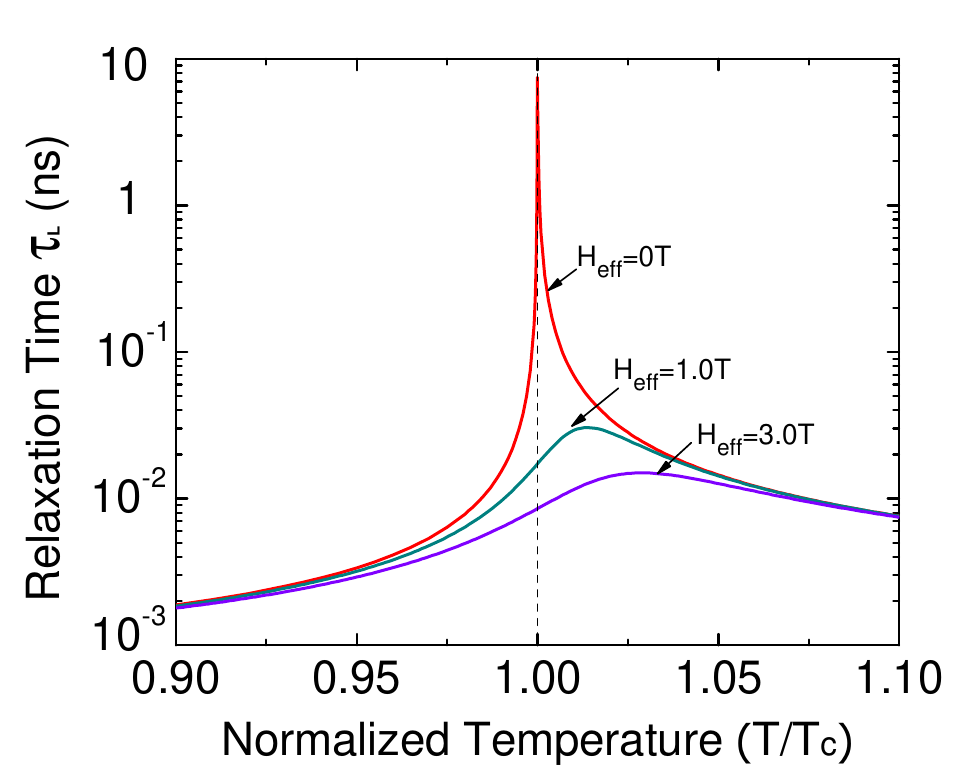}
  \caption{(Color online) The longitudinal relaxation time $\tau_l$ as a function of temperature
  for several magnetic fields. We choose a small difference between $m$ and
  $m_{\rm eq}$ at $t=0$ and identify $t=\tau_l$ where the difference
  is reduced by the half. We have used $\tau_s =1$ ps and
  $S=1/2$.}
\end{figure}

\begin{figure}
  \centering
  \includegraphics[width=8.5cm]{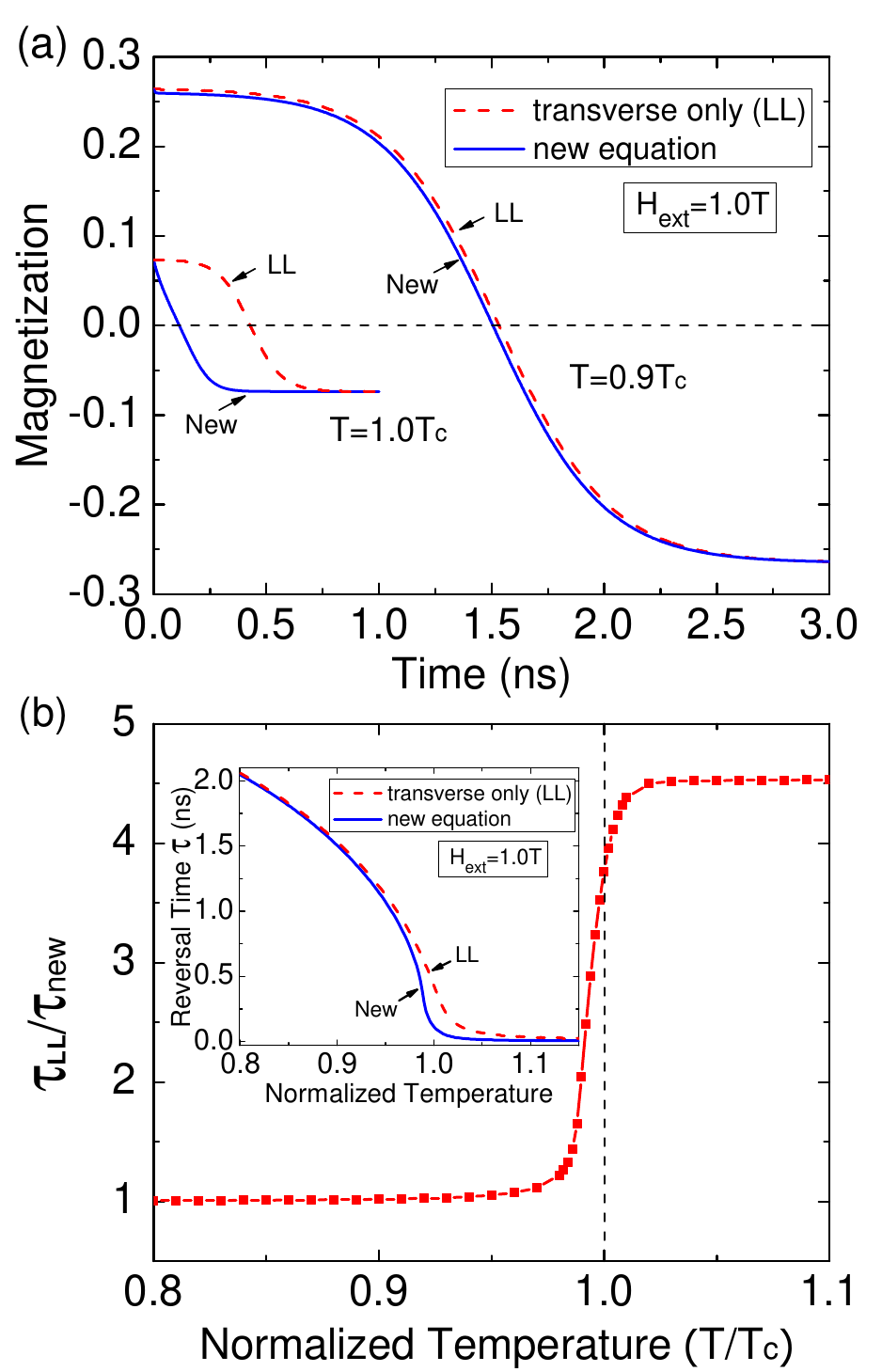}
  \caption{(Color online) a) The time dependence of magnetization reversal when a reversal
  magnetic field is applied at $t>0$ for the temperature $T=0.9T_c$ and $T=1.0T_c$ as indicated. The
  blue (solid) and red (dashed) curves were obtained by Eq.~(6) and by the first two terms of
  Eq.~(9), respectively. b)
  the reversal times and their ratio as a function of the temperature obtained from a). The
  parameters are $\tau_s=1.0$ ps, $S=1/2$,
  the anisotropy constant $K=0$, and the external field $H=1.0$ T.}
\end{figure}

In order to gain more quantitative insight for the interplay between
the transverse and longitudinal relaxations, we consider several
simple cases where the numerical calculations can be readily
performed. We assume that the magnetic particle is a single domain
so that there is no spatial dependence of the effective field and
the magnetization. Furthermore, the long-range magnetostatic field
is also discarded. In the first case, we compare the reversal times
with and without the longitudinal relaxation in a simplest case: an
isotropic magnetic particle (zero magnetic anisotropy) is initially
magnetized at $5^\circ$ from $+z$ axis and a reversal magnetic field
in the direction of $-z$ is applied at $t>0$. Figure 2(a) shows the
importance of the longitudinal relaxation when the temperature
approaches Curie temperature. We compare the magnetization dynamics
with and without the last term of Eq.~(9). If the temperature is
considerably below the Curie temperature, e.g., $T=0.9T_c$, the
longitudinal relaxation term has a negligible effect, i.e., the
result is essentially same whether the last term of Eq.~(9) is
included. This is because the magnitude of the magnetization is
nearly time-independent at low temperature. When the temperature is
near the Curie temperature, the magnitude of the magnetization is
significantly reduced. More importantly, the magnitude is now a
function of time due to its dependence on the total effective field.
In this case, there is a much difference if one includes the
longitudinal relaxation. In Fig.~2(b), we show the ratio of the
reversal times calculated with and without the longitudinal
relaxation. Clearly, the reversal time from Eq.~(6) is much faster
than that of the LL equation if the temperature is close to or
higher than Curie temperature.

Next we apply our equation to a hypothetical HAMR process when the
laser heating and thermal diffusion produce a time-dependent
temperature profile: the temperature of the particle increases
linearly  $T(t) = T_{\rm rm} +(t/t_{\rm heat}) (T_{\rm p} -T_{\rm
rm})$ from the room temperature $T_{\rm rm}$ to a peak value $T_{\rm
p}$ for the period of $0<t<t_{\rm heat}$ of lasing application.
After the heating process is completed and the laser is removed, the
temperature decreases due to heat diffusion into surroundings. We
assume the temperature is $T(t)=T_{\rm rm}+(T_{\rm p} -T_{\rm rm})
\exp[-(t-t_{\rm heat})/t_{\rm cool}]$ for $t>t_{\rm heat}$. While
the precise temperature profile should be determined via heat
transport equations with proper boundary conditions, our
hypothetical temperature is characterized by three parameters: the
peak temperature of the particle $T_{\rm p}$, and the heating and
cooling rates $1/t_{\rm heat}$ and $1/t_{\rm cool}$. We choose the
low-temperature magnetic anisotropy field much larger than the
external magnetic field so that the magnetic reversal does not occur
at the room temperatures. The temperature dependence of the
anisotropy energy $E_a$ is modeled by $E_a =K m^2(T) \sin^2\theta$,
where $m(T)$ is the magnitude of the magnetization at temperature
$T$ and $\theta$ is the angle between the magnetization vector and
$z$-axis \cite{Thiele,Nowak}. By placing the above temperature
profile and effective magnetic fields into Eq.~(6), we have
numerically calculated the time dependent magnetization shown in
Fig.~(3). As we expected, the magnetization reversal requires a high
peak temperature $T_{\rm p}$ to reduce the anisotropy. The rates of
heating and cooling are also important; they should be slow enough
so that the magnetization has sufficient time to relax to the ground
state via transverse and longitudinal relaxations.

\begin{figure}
 \centering
  \includegraphics[width=9cm]{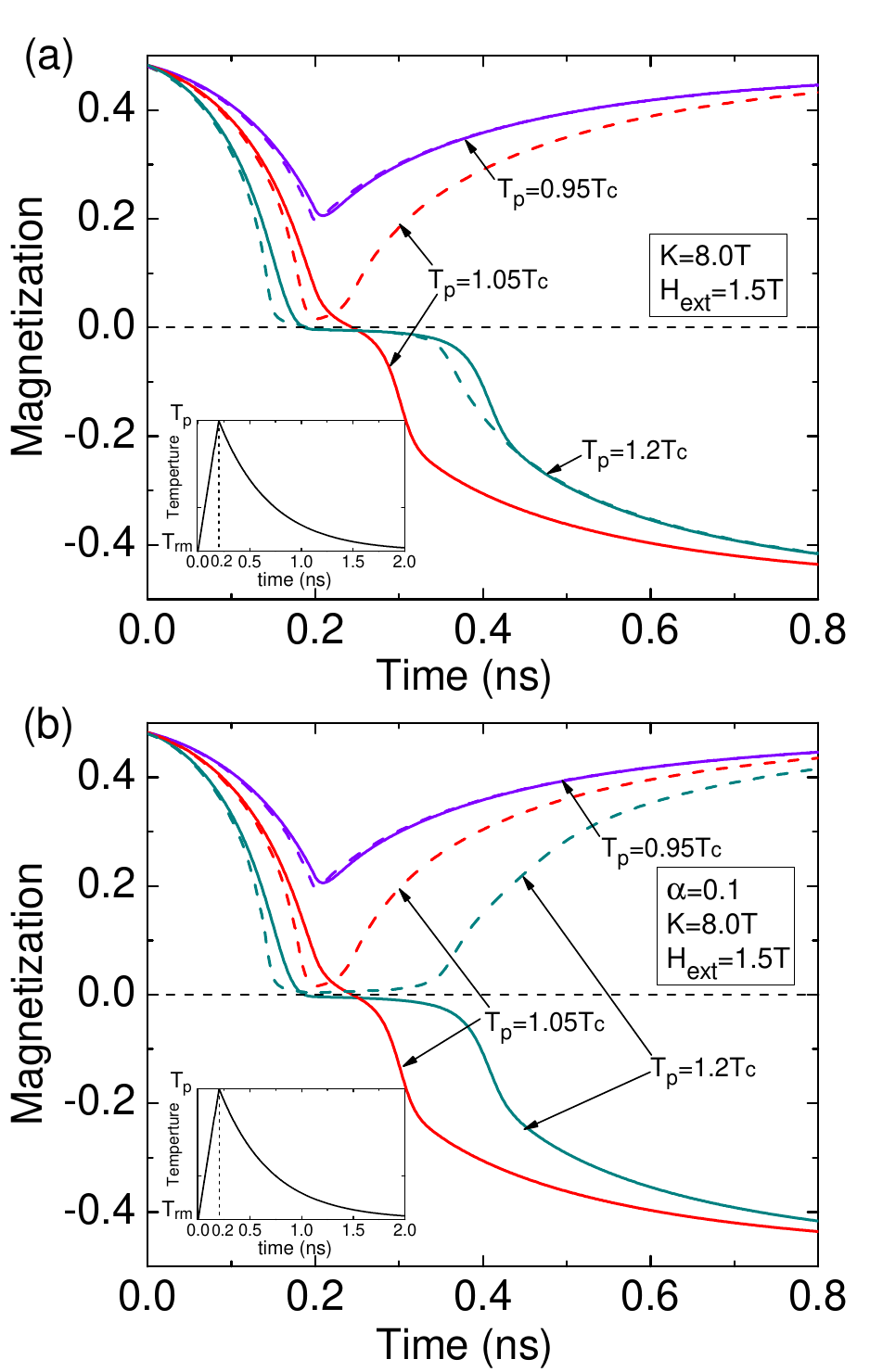}
  \caption{(Color online) Time dependence of magnetization for given
  temperature profiles after a reversal magnetic field $H=1.5$T is applied
  for $t>0$. a) The results are obtained from Eq.~(6) (solid curves) and from
  Eq.~(15) (dashed curves). b) Comparison of the results from Eq.~(6) and from the conventional LL with
  a constant damping parameter (dashed curves). The Inserts are the
  hypothetical temperature profiles.
  The parameters are $K=8.0$T, $S=1/2$, and $\tau_s =1.0$ ps.}
\end{figure}

More quantitatively, we have made two comparisons in Fig.~3. First,
we compare our equation with a modified LL which allows the
magnitude of the magnetization varying with the time due to changing
temperature in HAMR,
\begin{equation}
\frac{d{\bf m}}{dt} = \frac{d(m_{\rm eq} \hat{\bf m})}{dt} = -\gamma
{\bf m}\times {\bf H}_{\rm eff} - \gamma \frac{\alpha_{\rm tr}}{m}
{\bf m}\times ({\bf m}\times {\bf H}_{\rm eff}) + \frac{dm_{\rm
eq}}{dT} \cdot \frac{dT}{dt} \hat{\bf m}
\end{equation}
where the transverse damping parameter is given by Eq.~(10). The
above equation implies that the magnitude of the magnetization is
always in equilibrium with the instantaneous temperature, i.e., the
longitudinal relaxation is infinite fast. Fig.~3(a) shows that such
approximation is quite accurate even for the temperature
$T_p=0.95T_c$. However, the deviation begins to show up when the
peak temperature is higher than the Curie temperature. In Fig.~3(b),
we further compare our results with a constant damping parameter
(i.e., taking $\alpha_{\rm tr}$ in Eq.~(15) as a constant). The
deviation of this conventional LL with ours becomes more significant
at high temperatures. For example, even for $T_p =1.2 T_c$, the
magnetization reversal is not possible from the conventional LL
equation, see Fig.~3(b).

To end this section, we should briefly compare our equation with the
LLB equation of Garanin \cite{Garanin}. Since the LLB equation is
based on the atomistic LL equation that we believe is questionable,
we should not make extensive comparisons. We point out that the LLB
equation also contains the transverse and longitudinal relaxations,
and the essential difference is the temperature dependence of the
relaxation parameters. At low temperatures, both our equation and
the LLB equation reduce to the conventional LL equation. At high
temperatures, the relaxations in the LLB equation depend explicitly
on temperatures; this is because the longitudinal relaxation to the
equilibrium magnetization is solely controlled by the classical
random field which is proportional to the temperature. In our case,
the dependence of the relaxation time on temperature is implicit,
via the temperature dependence of the equilibrium magnetization.
More importantly, the instantaneous relaxation time $\tau_s$ in our
equation has microscopical meaning as the scattering lifetime of
electron spins while the damping parameter in the atomistic LL does
not have a microscopic counterpart.

\section{Stochastic fields}

Our proposed equation, Eq.~(6), describes the time-dependence of the
average magnetization. The fluctuation at the finite temperature,
particularly at a high temperature, becomes important. To address
the fluctuation, one should include stochastic fields in the
macroscopic dynamic equation. Similar to Brown's method \cite{Brown}
for the LL equation, we introduce the stochastic fields ${\bf h}
(t)$ as follows,
\begin{equation}
\frac{d{\bf m}}{dt} = -\gamma {\bf m}\times ({\bf H}_{\rm eff} +
{\bf h}) - \frac{{\bf m} -{\bf m}_{\rm eq}}{\tau_s} .
\end{equation}
We point out that the stochastic field does not enter in the
relaxation term although the instantaneous equilibrium magnetization
depends on the total field ${\bf H}_t$. The reason is as follows.
The interaction between the random field and the magnetization is
$-{\bf m}\cdot {\bf h}(t)$. This interaction gives arise a random
torque on the magnetization $-\gamma {\bf m}\times {\bf h}(t)$ that is
added to the deterministic torque equation. As in the case of the
Brownian motion, the Langevin random field ${\bf f}(t)$ is only
included in the particle motion $d^2 r/dt^2 = -\alpha {\bf v} + {\bf
F}(t) + {\bf f}(t)$ (where the friction force $-\alpha {\bf v}$ and
the external driven force ${\bf F}(t)$ are not changed by the random
force). To determine the magnitude and the correlation of the
stochastic fields, we first write the above stochastic equation in
the standard form of Langevin,
\begin{eqnarray}
\frac{dm_i}{dt} = \left [ -\gamma {\bf m}\times {\bf H}_{\rm eff} - \frac{{\bf m} -{\bf m}_{\rm eq}}{\tau_s} \right ]_i - \gamma \sum_{jk} {\varepsilon}_{ijk} m_j h_k.
\end{eqnarray}
where ${\varepsilon}_{ijk}$ is the Levi-Civita symbol. The corresponding Fokker-Planck equation is thus have the
following form,
\begin{equation}
\frac{\partial P}{\partial t} = -\sum_{i} \frac{\partial }{\partial m_i} \left [ \left ( -\gamma {\bf m}\times {\bf H}_{\rm eff} - \frac{{\bf m} -{\bf m}_{\rm eq}}{\tau_s} \right )_i + D \gamma^2 {\bf m}\times \left({\bf m}\times \frac{\partial}{\partial {\bf m}}\right)_i \right ] P
\end{equation}
where $P$ is the probability density and $D$ is the random field
correlation constant. At the equilibrium, one may assume that the
probability density takes a simple Boltzmann distribution, i.e.,
$P\propto \exp(-{\bf m}\cdot {\bf H})$. By placing this form of $P$
into Eq.~(18), one finds the desired correlation of the random field
given below,
\begin{equation}
< h_i (t) h_j (0) >  = \frac{2 k_B T \alpha_{\rm tr}}{\gamma m V} \delta_{ij} \delta (t)
\end{equation}

\section{Summary}

In this paper, we have proposed a model of magnetization dynamics
for an entire range of temperature based on the quantum kinetic
approach with the instantaneous local relaxation time approximation.
The resulting equation generates a low temperature magnetization
dynamic same as the Landau-Lifshitz equation, namely, the transverse
magnetization is sufficient to describe dynamics. When the
temperature approaches or exceeds the Curie temperature, it is
essential to include the longitudinal magnetization relaxation. With
our new dynamic equation, one can model the entire heat-assisted
magnetic recording processes when the temperature are heated and
cooled through the Curie temperature \cite{Torabi,Bunce,Mercer}. The
stochastic fields on the magnetization are also proposed. This work
is partially supported by the U.S. DOE (DE-FG02-06ER46307) and by
the NSF (ECCS-1127751).

\bibliography{apssamp}

\end{document}